\documentclass[aps,prl,twocolumn,groupedaddress]{revtex4}

\usepackage{graphicx}

\begin{document}

\title{Wave Function Collapse in a Mesoscopic Device}

\author{G.B.\ Lesovik$^{a}$, A.V.\ Lebedev$^{a}$, and G.\ Blatter$^{b}$}

\affiliation{$^{a}$L.D.\ Landau Institute for Theoretical Physics RAS,
117940 Moscow, Russia}

\affiliation{$^{b}$Theoretische Physik, ETH-H\"onggerberg, CH-8093
Z\"urich, Switzerland}

\date{\today}

\begin{abstract}
   We determine the non-local in time and space current-current
   cross correlator $\langle \hat{I}(x_1,t_1) \hat{I}(x_2,t_2)
   \rangle$ in a mesoscopic conductor with a scattering center
   at the origin. Its excess part appearing at finite voltage
   exhibits a unique dependence on the retarded variable
   $t_1-t_2-(|x_1|-|x_2|)/ v_{\rm\scriptscriptstyle F}$, with
   $v_{\rm\scriptscriptstyle F}$ the Fermi velocity. The
   non-monotonic dependence of the retardation on $x_1$ and
   its absence at the symmetric position $x_1 = -x_2$ is a
   signature of  an instantaneous wave function collapse, which
   thus becomes amenable to observation in a mesoscopic solid
   state device.
\end{abstract}

\maketitle

The recent years have seen a confluence of interests in
quantum optics and condensed matter physics. This trend is
particularly apparent in the field of quantum information
science \cite{steane_98}, where quantum optical as well as
mesoscopic nanoscale devices are being designed and
implemented as potential hardware components for quantum
computing. Besides this technological aspect, fundamental
questions traditionally investigated in quantum optical
setups \cite{aspect_99} are now being implemented in
mesoscopic structures. Examples are the recent proposals
for solid state entanglers \cite{ssent} and their potential
use in testing Bell inequalities \cite{ssbell,beenakker_03}
or the fermionic implementation \cite{HBT} of
Hanburry-Brown-Twiss type experiments testing for particle
correlations induced by their statistical properties.
Another fundamental issue is the measurement process and the
associated collapse of the wave function. The latter has
lately been discussed in the context of quantum measurement
of quantum bits \cite{makhlin_01,korotkov_01,averin_02} with
the main focus on issues related to the back-action
dephasing of the qubit and the aquisition of information
by the detector. In the present paper, we concentrate
on the wave function collapse itself and demonstrate
how it can be identified and analyzed in a measurement
of current cross-correlators in a mesoscopic device.
\begin{figure}[h]
  \includegraphics[scale=0.44]{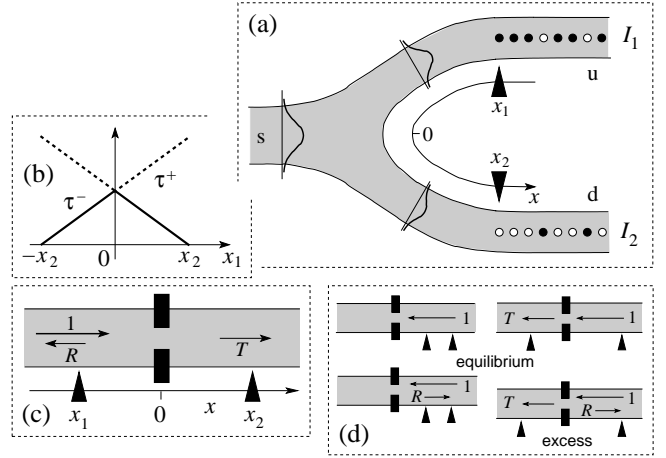}
  \caption[]{(a) A splitter directs an incident wave
  into arms `u' and `d' with amplitudes $t_\mathrm{su}$
  and $t_\mathrm{sd}$. After measurement in `u' (at
  $t_1=t$ and $x_1 < 0$) the wave is projected
  resulting in a specific sequence of particles. The
  simultaneous measurement in arm `d' (at $t_2 = t^+$
  and $x_2=-x_1>0$) will detect the conjugate sequence.
  Shifting the measurement point in `u' to smaller values $x_1$
  produces a delay $\tau^- =(|x_1|-x_2)/v_{\rm\scriptscriptstyle
  F}$ in the excess noise with the symmetric dependence shown
  in (b). The actual calculation of the current excess noise
  is carried out for the quantum wire with a scatterer
  characterized by transmission and reflection amplitudes $t$
  and $r$, see (c). The processes contributing to the
  equilibrium and excess noise are sketched in (d).}
  \label{fig:setup}
\end{figure}

In the orthodox interpretation of quantum mechanics, the wave
packet reduction is introduced as an independent postulate
within the context of the measurement process \cite{vonNeumann}.
While the ordinary time evolution of a quantum system follows the
dynamics described by the Schr\"odinger equation, the measurement
process involves an instantaneous projection onto the pointer
basis of the measurement device. Attempts to bind the wave
function collapse into the conventional frame of unitary time
evolution have been made, particularly in model systems
describing a quantum degree of freedom coupled to a reservoir
\cite{leggett_87}, but with limited success so far. The
experiment suggested and analyzed below will be suitable
to separate an instantaneous collapse from one carrying its
own dynamics through the measurement of retardation effects.

According to usual expectations, the detection of an individual
particle induces a wave function collapse, however, no useful
quantitative information on the collapse itself can be extracted
from such an isolated measurement. On the other hand, the wave
packet reduction appears naturally in von Neumann's prescription
of repeated measurements \cite{vonNeumann}, motivating its
experimental observation through {\it repeated} detection.  In
today's context this is realized in the measurement of correlators,
e.g., the current-current correlators (noise) $\langle \hat{I}(t)
\hat{I}(0)\rangle$ in a mesoscopic device. In such an experiment,
the second measurement tests the change in state induced by
the first measurement and hence carries the signature of
the wave function collapse.

In our theoretical analysis below we stay within the
framework set by the orthodox interpretation of quantum
mechanics. We determine the current cross correlator within
the second quantized formalism which treats the wave function
collapse as an instantaneous and nonlocal process. Accordingly,
our result carries the signature of an instantaneous collapse
as expressed through a vanishing delay time between the
appearance of particles (electrons) in one place and
the vanishing of their quantum-alternative partners
(appearance of holes) in the other place. On the other hand,
one expects that a collapse involving its own dynamics, e.g.,
the unitary Schr\"odinger evolution, naturally leads to a
finite delay which will show up in the noise experiment.
Hence the proposed experiment provides quantitative information
on the properties of the wave function collapse in a mesoscopic
device.

To fix ideas, consider a particle wave incident from a source lead
`s' and split with amplitudes $t_\mathrm{su}$ and $t_\mathrm{sd}$
into the upper (`u') and lower (`d') arms of a fork device,
see Fig.\ \ref{fig:setup}(a). We emphasize that it is the
unitary evolution dictated by wave mechanics which determines
the propagation of particles into the two arms. For the time
being, we ignore the possibility that the splitter projects
the particles and distributes them into the two arms via a classical
random process, i.e., we assume that there is no object in the
splitter associated with a local hidden variable. This assumption
has to be checked in the experiment and we will return back to this
point later.

Hence before any measurement, the particles propagate in
terms of waves and are delocalized between the two arms. A
measurement of the current in one of the arms, say `u',
projects the wave function and the subsequent evolution is
in terms of particle streams, see Fig.\ \ref{fig:setup}(a).
Note that we only need one measurement to project the waves
to particle streams in {\it both} arms. However, in order
to detect the (instantaneous) collapse of the wave function,
we have to perform two measurements in the two arms allowing us
to observe the coincidence between a particle missing in one arm and
the additional particle propagating in the other arm. Such an
experiment can be realized efficiently if detectors are used
which react on the presence of particles in one arm and
holes in the other arm. The observation of a perfect
coincidence between the appearance of particles and their
partner holes then is a demonstration of the instantaneous
reduction of the wave function in this setup.

The information that can be extracted from the noise
experiment depends crucially on its time resolution.
E.g., one may deliberately separate (in time) the stream of
coincident events, i.e., particles and holes in `d' and `u',
by choosing an asymmetric splitter with a small transmission
$T_\mathrm{sd} =|t_\mathrm{sd}|^2$ into one of the arms; this
type of splitter has been introduced by Beenakker {\it et al.}
\cite{beenakker_03} in a recent proposal for the measurement
of the degree of entanglement in a many body wave function.
A detector with limited temporal and/or spatial resolution
then is still capable of detecting individual events and thus
can serve in this type of coincidence experiment; however,
the limited resolution restricts the analysis of the wave
function collapse and its intrinsic dynamics. On the other
hand, if detectors with high resolution are used in the
measurement of cross-correlators (of either current or
density) a finite frequency/short time measurement can
trace the signature of the wave function collapse for
any value of the transmission $|t_\mathrm{su}|^2$.
Furthermore, a high resolution provides quantitative
details on the collapse itself; in particular, it allows
to determine accurately the delay time involved in the
collapse and hence an instantaneous collapse can be
distinguished from a dynamical one. Typical parameters
used in mesoscopic setups involve time scales of order
GHz and length scales of order micrometers --- one then
easily checks that a dynamical collapse involving
the Fermi velocity and beyond can be resolved, while
a dynamical collapse involving a (super)luminal velocity
is beyond the attainable resolution.

The above idea for a direct measurement of the wave function collapse
can be implemented with different types of mesoscopic experiments: a) a
beam splitter in a fork geometry can be realized with the help of
electrostatic gates structuring a two-dimensional electron gas as
done in Ref.\ \onlinecite{HBT}, b) a nearly ideal splitter can be
realized in a quantum Hall setup with a split gate \cite{HBT}, and
c) use can be made of a simple quantum wire with a localized
scatterer where the two arms correspond to the backward and
forward scattering channels, see Fig.\ \ref{fig:setup}(c). First,
we concentrate on the last example \cite{foot}, the quantum wire,
and determine the irreducible current-current cross-correlator
\begin{equation}
   C_{x_1,x_2}(t_1 -t_2) \equiv
  \langle\langle\hat I(x_1,t_1)\hat I(x_2,t_2)\rangle\rangle,
   \label{II}
\end{equation}
with the signal measured once on the same side of the scatterer
($x_1 x_2 >0$) and subsequently on opposite sides ($x_1 x_2< 0$).
In the coherent conductor studied here, the excess noise
$C^\mathrm{ex}_{x_1,x_2}(\tau;V) \equiv C_{x_1,x_2}(\tau;V)-
C_{x_1,x_2}(\tau;V=0)$ is entirely due to the quantum shot noise;
the latter has been intensely studied during the past years
\cite{shot_noise}. Most of these studies have concentrated
on the low-frequency limit, identifying quasi-particle charges
\cite{qp_charge} or anti-bunching of fermions \cite{HBT}, to name
two well-known examples. While the partitioning of the particle
beam due to the reduction of wave packets was clearly identified
as the source of shot noise \cite{shot_noise} this aspect has
never been analyzed in detail. The most interesting result
is found for the measurement involving current fluctuations
on opposite sides of the barrier: we find that the excess noise
$C_{x_1,x_2}^\mathrm{ex} (\tau)$ depends on a spatially retarded
variable with the particular form $\tau-\tau^-$ where $\tau^-
= (|x_1|-|x_2|)/ v_{\rm \scriptscriptstyle F}$, see Fig.\
\ref{fig:setup}(b). This should be contrasted with the
ballistic retardation appearing in the equilibrium noise
$C^\mathrm{eq}_{x_1,x_2} (\tau;V=0)$ and exhibiting the causally
retarded dependence $\tau-\tau^+$ with $\tau^+ =
(|x_1|+|x_2|)/ v_{\rm\scriptscriptstyle F}$ involving the
ratio of the travelling distance and the Fermi velocity. This
latter type of retardation has to be expected due to the relation
between the equilibrium correlator and the
(causally retarded) linear response function enforced by
the fluctuation-dissipation theorem. On the contrary, the
particular dependence on $\tau-\tau^-$ appearing in
$C^\mathrm{ex}$ identifies the presence of instantaneous
correlations between spatially separated events, which we
interpret as arising from the instantaneous collapse of the
wave function.

We now turn to the derivation of the above results and determine
the excess noise in the current-current cross-correlator. We
concentrate on the geometry sketched in Fig.\ \ref{fig:setup}(c)
and define the field operators (for one spin component;
$v_\epsilon= \sqrt{2m\epsilon}$)
\begin{eqnarray}
   &&\hat \Psi|_{x<0}
   =\!\int\!\! \frac{d\epsilon}{\sqrt{h v_\epsilon}}
      \bigl[(e^{ikx}\!+r_\epsilon e^{-ikx})\,\hat a_\epsilon\!
     +t_\epsilon e^{-ikx} \hat b_\epsilon\bigr]
     e^{-\frac{i\epsilon t}{\hbar}}
   \nonumber \\
   &&\hat\Psi|_{x>0}
   =\!\int\!\! \frac{d\epsilon}{\sqrt{h v_\epsilon}}
      \bigl[t_\epsilon e^{ikx} \hat a_\epsilon +(e^{-ikx}
     +r^\prime_\epsilon e^{ikx}) \hat b_\epsilon\bigr]
     e^{-\frac{i\epsilon t}{\hbar}}
   \nonumber
\end{eqnarray}
with $\hat a_\epsilon$ ($\hat b_\epsilon$) the electronic
annihilation operators for the left (right) reservoir and $t$,
$r$, $r^\prime$ the usual scattering amplitudes. Substituting
these expressions into the current operator $\hat I(x,t) =
({ie\hbar}/{2m}) [\partial_x\hat\Psi^+(x) \hat\Psi(x)-\hat
\Psi^+(x)\partial_x\hat\Psi(x)]$ and using the standard
scattering theory of noise \cite{lesovik_89,shot_noise,lesovik_99},
we obtain the expression for the current-current cross
correlator (\ref{II}). We split the result into an
equilibrium part $C_{x_1,x_2}^\mathrm{eq}(\tau)$ and an excess
part $C_{x_1,x_2}^\mathrm{ex}(\tau)$ with $\tau = t_1-t_2$;
correlators evaluated at the same side of the scatterer are
denoted by $\bar{C}$, those on opposite sides by $C$. Assuming
$|\epsilon^\prime-\epsilon| \sim eV \ll \epsilon_{\rm
\scriptscriptstyle F}$, with $V$ the applied voltage and
$\epsilon_{\rm \scriptscriptstyle F}$ the Fermi energy, we drop
terms \cite{lesovik_99} small in the parameter
$|\epsilon^\prime-\epsilon|/\epsilon_{\rm\scriptscriptstyle F}$
and find the result for $x_1 x_2 <0$ (the Fermi occupation
numbers $n_{\rm\scriptscriptstyle L}(\epsilon)$ and
$n_{\rm \scriptscriptstyle R}(\epsilon)$ denote the
filling of the attached reservoirs),
\begin{eqnarray}
      &&C_{x_1,x_2}^\mathrm{eq}(\tau)
      =\frac{2e^2}{h^2}\int d\epsilon \, d\epsilon^\prime
      e^{i(\epsilon^\prime-\epsilon)\tau/\hbar}
      \label{C_eq_pm} \\
      &&\times \bigl[t_{\epsilon^\prime}t^*_{\epsilon}
      e^{i(\epsilon^\prime-\epsilon)\tau^+/\hbar}
      \,n_{\rm\scriptscriptstyle L}(\epsilon^\prime)
      [1-n_{\rm\scriptscriptstyle L}(\epsilon)]
      \nonumber\\
      &&+t^*_{\epsilon^\prime}t_{\epsilon}
      e^{-i(\epsilon^\prime-\epsilon)\tau^+/\hbar}
      \,n_{\rm\scriptscriptstyle R}(\epsilon^\prime)
      [1-n_{\rm\scriptscriptstyle R}(\epsilon)]\bigr],
      \nonumber
\end{eqnarray}
while the corresponding result evaluated on the same side of the
scatterer ($x_1 x_2 >0$) takes the form
\begin{eqnarray}
      &&\bar{C}_{x_1,x_2}^\mathrm{eq}(\tau)
      =\frac{2e^2}{h^2}\int d\epsilon\, d\epsilon^\prime\,
      e^{i(\epsilon^\prime-\epsilon)\tau/\hbar}
      \label{C_eq_pp}\\
      && \times \bigl[ T_{\epsilon^\prime}
      e^{-i(\epsilon^\prime-\epsilon)\tau^-/\hbar}
      n_{\rm\scriptscriptstyle L}(\epsilon^\prime)
      [1-n_{\rm\scriptscriptstyle L}(\epsilon)]
      \nonumber\\
      &&+(e^{i(\epsilon^\prime-\epsilon)\tau^-/\hbar}
      +R_{\epsilon^\prime}
      e^{-i(\epsilon^\prime-\epsilon)\tau^-/\hbar})
      \,n_{\rm\scriptscriptstyle R}(\epsilon^\prime)
      [1-n_{\rm\scriptscriptstyle R}(\epsilon)]
      \nonumber\\
      &&-(r^\prime_{\epsilon^\prime}r^{\,\prime*}_{\epsilon}
      e^{i(\epsilon^\prime-\epsilon)\tau^+/\hbar}
      +c.c.\,)\,
      n_{\rm\scriptscriptstyle R}(\epsilon^\prime)
      [1-n_{\rm\scriptscriptstyle R}(\epsilon)]\bigr].
      \nonumber
\end{eqnarray}
The time dependence appearing in (\ref{C_eq_pm}) and
(\ref{C_eq_pp}) involves the retardations
\begin{equation}
   \tau^\pm = (|x_1|\pm |x_2|)/v_{\rm \scriptscriptstyle F}
   \label{tau_pm}
\end{equation}
with $v_{\rm\scriptscriptstyle F}$ the Fermi velocity. The excess
part $C_{x_1,x_2}^\mathrm{ex}(\tau)$ is given by the expressions
\begin{eqnarray}
      &&C_{x_1,x_2}^\mathrm{ex}(\tau)
      =\frac{2e^2}{h^2}\int d\epsilon \, d\epsilon^\prime
      e^{i(\epsilon^\prime-\epsilon)(\tau-\tau^-)/\hbar}
      \label{C_ex_pm} \\
      &&\times\>
      t_{\epsilon}^*t_{\epsilon^\prime}
      r^*_{\epsilon^\prime} r_{\epsilon}
      \,[n_{\rm\scriptscriptstyle L}(\epsilon^\prime)
      -n_{\rm\scriptscriptstyle R}(\epsilon^\prime)]
      \,[n_{\rm\scriptscriptstyle L}(\epsilon)
      -n_{\rm\scriptscriptstyle R}(\epsilon)],
      \nonumber\\
      &&\bar{C}_{x_1,x_2}^\mathrm{ex}(\tau)
      =\frac{2e^2}{h^2}\int d\epsilon \, d\epsilon^\prime
      e^{i(\epsilon^\prime-\epsilon)(\tau-\tau^-)/\hbar}
      \label{C_ex_pp}\\
      &&\times
      [T_{\epsilon^\prime}R_{\epsilon}\,
      n_{\rm\scriptscriptstyle L}(\epsilon^\prime)-
      R_{\epsilon^\prime}T_{\epsilon}\,
      n_{\rm\scriptscriptstyle R}(\epsilon^\prime)]
      [n_{\rm\scriptscriptstyle L}(\epsilon)-
      n_{\rm\scriptscriptstyle R}(\epsilon)],
      \nonumber
\end{eqnarray}
with the unique retardation $\tau^-$. In the following, we drop
the energy dependencies of the scattering amplitudes, allowing us
to perform the integration over energies, and we find the
simplified expressions (we denote the temperature by $\theta$ and
assume $k_{\rm\scriptscriptstyle B} = 1$)
\begin{eqnarray}
    &&C_{x_1,x_2}^\mathrm{eq}(\tau,\theta)
    =-\,\frac{2e^2T}{h^2}
    \bigl[\alpha(\tau\!+\!\tau^+,\theta)
         +\alpha(\tau\!-\!\tau^+,\theta)\bigr],
    \nonumber\\
    &&\bar{C}_{x_1,x_2}^\mathrm{eq}(\tau,\theta)
    =-\,\frac{2e^2}{h^2}
    \bigl[\alpha(\tau\!+\!\tau^-,\theta)
         +\alpha(\tau\!-\!\tau^-,\theta)
    \label{C_eqx_T} \\
    &&\qquad\qquad\qquad\quad -R[\alpha(\tau\!+\!\tau^+,\theta)
         +\alpha(\tau\!-\!\tau^+,\theta)]\bigr],
    \nonumber\\
    &&C_{x_1,x_2}^\mathrm{ex}(\tau,\theta)
    =\frac{8e^2TR}{h^2}
    \sin^2\!\!\left[\frac{eV(\tau\!-\!\tau^-)}{2\hbar}\right]\!
    \alpha(\tau-\tau^-\!,\theta),
    \nonumber
\end{eqnarray}
with the temperature dependence given by the expression
$\alpha(\tau, \theta) = \pi^2\theta^2/\sinh^2 [\pi\theta\tau
/\hbar]$; in the zero temperature limit this reduces to
$\alpha(\tau,0) = \hbar^2/\tau^2$. The singularity at $\tau
\rightarrow 0$ is cutoff for $\tau < \hbar/\epsilon_{\rm
\scriptscriptstyle F}$ and the equilibrium correlator
changes sign as individual fermions extending over the
Fermi wavelength $\lambda_{\rm\scriptscriptstyle F}$ are
probed; proper calculation of this feature requires to
account for the finite Fermi energy and bandwidth of the
electron system. Quite remarkably, the excess noise is
given by a unique expression and involves only the
retardation $\tau^-$. The above results apply for the
quantum wire, cf.\ Fig.\ \ref{fig:setup}(b). The result
(\ref{C_eqx_T}) for the excess noise is easily rewritten
for the fork geometry in Fig.\ \ref{fig:setup}(a) by
replacing the product of transmission and reflection
probabilities $TR$ by the product $-T_\mathrm{su}
T_\mathrm{sd}$, with $T_\mathrm{su}$ and $T_\mathrm{sd}$
the transmission probabilities from the soure lead
`s' into the upper (`u') and lower (`d') leads. The
sign change is due to the current reversal as the reflected
beam in the quantum wire is replaced by a second forward
directed beam in the fork geometry.

Let us analyze the results (\ref{C_eqx_T}) in more detail.
Consider first the {\it equilibrium noise}: The sign of
the correlator follows from the fact that a, say positive,
current fluctuation is followed by a compensating and
hence negative excursion. The terms
$\propto 1$ and $\propto R$ appearing in $\bar{C}^\mathrm{eq}$
derive from correlations in the incident flow and between the
incident and reflected flow, while $C^\mathrm{eq}$ measures
correlations between the incident and transmitted waves and hence
involves the transmission coefficient $T$, see the diagrams in
Fig.\ \ref{fig:setup}(d). The signs are as expected from the above
argument (note the sign change in the term $\propto R$ due to the
current reversal) and all retardations are causal involving the
geometric distance between particle detection. The symmetry $\tau
\leftrightarrow -\tau$ is due to the equivalence of the two
reservoirs injecting particles symmetrically under equilibrium
conditions.

The {\it excess noise} measures correlations between the
transmitted and reflected particles, see Fig.\ \ref{fig:setup}(d).
Its retardation and sign are those expected assuming an
instantaneous collapse of the wave function. I.e., projecting
the wave by the detection of an electron at $x_1$ implies
the instantaneous appearance of a hole at $x_2$ travelling
in the opposite direction, thus resulting in a positive sign
of $C^\mathrm{ex}$ (note the change in sign when going
from the point-contact to the fork geometry). Furthermore,
the vanishing of the relaxation time $\tau^-$ right at
the symmetric location $x_1 = -x_2$ is the hallmark of the
\textit{instantaneous} collapse of the wave function.
On the other hand, the observation of a nonzero time
delay (at the symmetric location $x_1 = -x_2$) would
indicate the presence of a non-trivial dynamical element in
the process of wave function collapse beyond the framework
of the orthodox theory with its projection postulate.
Hence measuring the excess noise in an experiment and
comparing to our result (\ref{C_eqx_T}) allows to confirm
or refute the instantaneous and nonlocal nature of the
wave function collapse. Finally, the oscillations
appearing in the excess noise are a consequence of the
sharp Fermi surfaces, their scale $\delta\tau \sim h/eV$
being determined by the voltage shift $eV$ between the
reservoirs; a temperature $\theta > eV$ smears this sharp
shift and the tails with their oscillations vanish
exponentially $\propto \exp(-2\pi\tau \theta/\hbar)$.

Above, we have emphasized the quantum nature of wave propagation
in our determination of the excess noise. One may ask about
the outcome of this experiment within a classical model of
electronic transport, where the splitter randomly distributes
the (ordered stream of) particles among the two arms of the
fork (see Fig.\ 1(a)), or, in our geometry, in the forward and
backward directions (see Fig.\ 1(c)). Indeed, particles sent into
the forward direction then are correlated with missing
particles (holes) in the backward flow and the correlator
has the same sign and retardation as in the quantum case.
Note, that the particular retardation given by $\tau^-$ has a
different origin in the classical and in the quantum case:
in the classical situation where particle-hole pairs are
{\it locally} generated at the splitter, the delay derives from
the difference in the travelling times of the particle and
the hole, while in the quantum case, the particle-hole pairs
appear due to the non-local process of {\it instantaneous}
projection.

A meaningful experiment has to distinguish between the
classical and the quantum mechanical scenario producing
the measured results. In order to show that quantum mechanics
is at work one has to confirm the wave propagation in the
device prior to measuring the current cross correlator.
This can be achieved through the observation of a
coherence phenomenon and we discuss two specific setups
in the following:

{\it i)} Inserting a second barrier, the observation of
resonant transmission through the interferometer formed
by the double barrier system confirms the wave propagation
in the device. An implementation using electrostatic
gates modulating a 2DEG allows to manipulate the second
barrier without significant perturbation of the remaining
sample.

\begin{figure}[h]
  \includegraphics[scale=0.44]{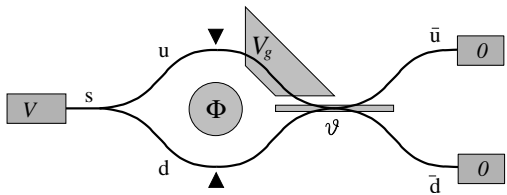}
  \caption[]{Experimental setup testing the wave propagation
  of electrons before projecting the wave function in the
  noise measurement. In a first step, the phase difference
  $\delta \varphi = \varphi_\mathrm{u}-\varphi_\mathrm{d}$
  picked up in the propagation along the upper (`u') or
  lower arm (`d') and the mixing angle $\vartheta$ of the
  four-beam splitter are arranged such that no electrons
  propagate into the upper arm `$\bar{\mathrm{u}}$';
  the phase $\delta \varphi$ may be manipulated through
  changing the flux $\Phi$ through the loop or via
  an electric gate potential $V_g$. Next, the noise
  correlator is measured (black triangles), projecting
  the electrons in the arms `u' and `d' and transforming
  the wave propagation into a flow of discrete particles.
  As a result of this projection, a finite current turns
  on in the upper arm `$\bar{\mathrm{u}}$'.}
  \label{fig:exp}
\end{figure}

{\it ii)} Following ideas developed within the context of the
famous double slit (Gedanken) experiment, we propose the
specific setup sketched in Fig.\ \ref{fig:exp} which tests
the particle-wave duality during the experiment. The incident
particle beam `s' is split into an upper (`u') and lower (`d')
arm (fork geometry) and subsequently recombined and redirected
into the leads `$\bar\mathrm{u}$' and `$\bar\mathrm{d}$' with the
help of a tunable reflectionless four-beam splitter. The phase
difference $\delta \varphi = \varphi_\mathrm{u} -
\varphi_\mathrm{d}$ picked up during the propagation in the
upper and lower leads can be tuned, either via a magnetic
flux $\Phi$ threading the loop or via an additional gate
electrode biasing (with $V_g$) one of the arms. The second
beam splitter is characterized through its transfer matrix
$M^{\bar{\mathrm{u}}\bar{\mathrm{d}}}_{\mathrm{u}\mathrm{d}}$,
\begin{equation}
   \left(\begin{array}{c}\bar\mathrm{u}\\
   \bar\mathrm{d}\end{array}\right)=
   \underbrace{\left(
   \begin{array}{cc}
   e^{i\phi}\cos\vartheta& -e^{i\psi}\sin\vartheta\\
   e^{-i\psi}\sin\vartheta& e^{-i\phi}\cos\vartheta
   \end{array}\right)}_{M^{\bar{\mathrm{u}}
   \bar{\mathrm{d}}}_{\mathrm{u}\mathrm{d}}}
   \left(\begin{array}{c}\mathrm{u}\\
   \mathrm{d}\end{array}\right),
   \label{M}
\end{equation}
with the angles $\vartheta\in(0,\pi/2)$, $\phi,\psi\in(0,2\pi)$;
without loss of generality we assume $\phi=\psi=0$. The wave function
behind the splitter then can be written in the form
\begin{eqnarray}
   \bar{\Psi}_{\bar{\mathrm{u}} \bar{\mathrm{d}}}
   &=& (\cos\vartheta\, e^{i\varphi_\mathrm{u}}\, t_\mathrm{su}
     -\sin\vartheta\, e^{i\varphi_\mathrm{d}}\, t_\mathrm{sd})
     |\bar{\mathrm{u}}\rangle\nonumber\\
   &&+ (\sin\vartheta\, e^{i\varphi_\mathrm{u}}\, t_\mathrm{su}
     +\cos\vartheta\, e^{i\varphi_\mathrm{d}}\, t_\mathrm{sd})
   |\bar{\mathrm{d}}\rangle.
   \label{Psiud}
\end{eqnarray}
The four-beam splitter shall be tuned such that all
electrons propagate to only one of the output leads, say
the down lead `$\bar\mathrm{d}$', implying the condition
$\tan\vartheta_0 =  \sqrt{T_\mathrm{su} /T_\mathrm{sd}}
\exp[i(\delta\varphi+\chi_\mathrm{su}-\chi_\mathrm{sd})]$,
where we have separated the amplitudes and phases of the
transmission coefficients, $t_\mathrm{su} = \sqrt{T_\mathrm{su}}
\exp(i\chi_\mathrm{su})$ and similar for $t_\mathrm{sd}$.
Tuning the phase $\delta\varphi+\chi_\mathrm{su}-
\chi_\mathrm{sd}$ to a multiple of $2\pi$ and choosing
the appropriate angle
\begin{equation}
   \tan\vartheta_0 = \sqrt{T_\mathrm{su} /T_\mathrm{sd}}
   \label{vartheta_0}
\end{equation}
one may redirect the recombined waves into the down lead.
Furthermore, subsequent scanning of the phase $\delta\varphi$
will produce an oscillating current in the output lead
`$\bar\mathrm{d}$', analogous to the intensity oscillations
observed on the detector screen in the double split experiment.
The observation of current oscillations as a function of
$\delta \varphi$ proves the coherent wave propagation
of the electrons through the device.

After establishing the quantum nature of the electron
propagation, the wave packet reduction is investigated
through a measurement of the current cross-correlator
between the two arms `u' and `d'. This measurement and
its concommitant wave function collapse will transform
the wave propagation in the two arms into streams of
particles. As a consequence of the projection through
the measurement a finite current will appear in the
upper lead `$\bar\mathrm{u}$'. The magnitude of this
current is determined with the help of the density matrix
$\bar{\rho}$ behind the four-beam splitter: transforming
the density matrix $\rho = T_\mathrm{su}|{\mathrm{u}}
\rangle\langle{\mathrm{u}}| +T_\mathrm{sd}|{\mathrm{d}}
\rangle\langle{\mathrm{d}}|$ describing the electron
streams in the two arms `u' and `d' after the projection
with the help of (\ref{M}), we obtain the current $\langle
\hat{I}_{\bar{\mathrm{u}}} \rangle = 2(e^2/h)V
\bar{\rho}_{\bar{\mathrm{u}} \bar{\mathrm{u}}}$ with
\begin{eqnarray}
   \bar{\rho}_{\bar{\mathrm{u}}\bar{\mathrm{u}}} &=& \langle
   \bar{\mathrm{u}}|\bar{\rho}|\bar{\mathrm{u}}\rangle
   = T_\mathrm{su} \cos^2 \vartheta_0 +
     T_\mathrm{sd} \sin^2 \vartheta_0\nonumber \\
   &=& 2 \frac{T_\mathrm{su}T_\mathrm{sd}}
   {T_\mathrm{su}+T_\mathrm{sd}},
   \label{barrho}
\end{eqnarray}
where we have made use of (\ref{vartheta_0}) in the last equation.
For a reflectionless splitter we have $T_\mathrm{su}+T_\mathrm{sd}
= 1$ and the final result for the current appearing in the lead
`$\bar\mathrm{u}$' after projection takes the form
\begin{equation}
   \langle \hat{I}_{\bar{\mathrm{u}}}
   \rangle = 4(e^2/h)V T_\mathrm{su}T_\mathrm{sd}.
   \label{cl_current}
\end{equation}
The maximum difference between the currents with and without
projection is obtained for the symmetric splitter with
$T_\mathrm{su} = T_\mathrm{sd} = 1/2$.

The above two-step procedure confirming the wave propagation
of the electrons before the measurement of the current cross
correlator excludes a classical interpretation of the
features showing up in the noise correlator; analyzing
the time delay $\tau^-$ in the excess noise then provides
the seeked for information on the wave function collapse.
In particular, the instantaneous collapse should manifest
itself through a zero time delay if a symmetric setup
with $x_1 = -x_2$ is chosen; on the other hand, one
expects that a collapse within the frame of unitary
time evolution produces a finite delay which the present
experiment is able to detect, provided the time resolution
of the noise measurement is adequate.

The test for the instantaneous wave function collapse
discussed here is related to the non-local properties
of quantum mechanics. The standard test demonstrating
the non-local nature of quantum mechanics is due to Bell
\cite{bell}. Bell inequality tests produce different
outcomes within a classical framework (based on local
hidden variables) and within a quantum mechanical
description. On the other hand, the measurement of
correlators, while producing interesting results on
fundamental issues of quantum mechanics, too, cannot
separate between the quantum mechanical and the classical
predictions. A prominent example is the measurement of
strangeness correlations in the $\mathrm{K}^0
\bar{\mathrm{K}}^0$ system: The decay of the Kaons
prevents one from carrying out Bell inequality tests.
Still, the observation of oscillations in the strangeness
correlation provides information on the entanglement in
the Kaon wave function \cite{gisin_01}. Nevertheless,
this type of oscillations can be generated within
the framework of a hidden variable theory, too. In the present
case, the measurement of the noise correlator provides
information on the wave function collapse, in particular
its dynamics. Again, the experiment itself cannot separate
between quantum mechanical and classical predictions.
In fact, a local hidden variable at the splitter could
emulate the shape of the time resolved correlator
including even the oscillations on the time scale
$\tau_V = h/eV$. One then may assume one of the
following two viewpoints: {\it a)} Accepting the
applicability of quantum mechanics one only needs
to rule out the presence of dephasing in the device;
the absence of dephasing is most simply confirmed
through the observation of the time oscillation $\propto
\sin^2(eV\,\delta\tau/2\hbar)$ in the correlator itself.
{\it b)} Those critical about the validity of quantum
mechanics first have to confirm the wave dynamics in
the device; the experiments {\it i)} and {\it ii)}
described above are designed to achieve this goal.

The current correlator is not directly measured in an experiment;
e.g., an old-fashoned Amp\`ere meter determines the angular
excursion of the pick-up loop. One then has to relate
the correlator of this classical meter variable to the correlator
of the quantum system \cite{lesovik_98}. For a linear detector,
one expects that the measured classical correlator can be
constructed from the quantum correlator through a linear
connection. Conventional wisdom tells that it is the symmetric
correlator that appears in an actual measurement \cite{LL}.
Indeed, a recent analysis carried out for an Amp\`ere meter
measuring local current-correlations shows that the main
term in the response is determined by the symmetrized
correlator $[C(\tau)+C(-\tau)]/2$ \cite{lesovik_98}; however,
additional small corrections appear involving the anti-symmetrized
correlator, too. The generalization to measurements at spatially
separated locations relates the measured correlator to the fully
symmetrized expression $[C_{x_1,x_2}(\tau)+C_{x_1,x_2}(-\tau)
+C_{x_2,x_1}(\tau)+C_{x_2,x_1}(-\tau)]/4$ as the main term.

Alternatively, the signature of the wave function collapse may be
detected in a frequency domain experiment; the result for the
spectral power $S_{x_1,x_2}^\mathrm{ex}(\omega)=\int d\tau
C_{x_1,x_2}^\mathrm{ex}(-\tau)\,\exp(i\omega\tau)$
of the excess noise takes the form
\begin{eqnarray}
    &&S_{x_1,x_2}^\mathrm{ex}(\omega)
    =\frac{2e^2TR}{h}\,e^{-i\omega\tau^-}
    \Bigl[\frac{\hbar\omega+eV}{1-e^{-(\hbar\omega+eV)/\theta}}
    \label{S_ex}\\
    &&\qquad\qquad\qquad\qquad+\frac{\hbar\omega-eV}
    {1-e^{-(\hbar\omega-eV)/\theta}}-
    \frac{2\hbar\omega}{1-e^{-\hbar\omega/\theta}}\Bigr]
   \nonumber
\end{eqnarray}
and contains the characteristic delay $\tau^-$ as a phase factor.
Again, a detailed analysis is required in order to relate the
experimentally measured quantity to the current excess noise.
For the case of an inductive measurement with an $LC$-circuit and
coinciding positions $x_1=x_2$ such a study has been carried out
\cite{lesovikloosen_97}; here, it is the symmetrized
power $[S_{x_1,x_2}^\mathrm{ex}+S_{x_2,x_1}^\mathrm{ex}]_{\Omega>0}/2$
which can be measured via the charge fluctuations
on the capacitor in an $LC$-circuit with two pick-up loops at
$x_1$ and $x_2$.

The above analysis has been carried out within a non-interacting
approximation. Accounting for the effect of Coulomb interaction one
may worry that the noise signal is damped due to the smoothing
produced by (longitudinal) screening; the latter involves
the Fermi velocity $v_F$. On the other hand, the incoming
electrons propagate (also with Fermi velocity) in the form
of a regular sequence of wave packets separated by the single
particle correlation time \cite{lesoviklevitov} $\tau_V=h/eV$,
a consequence of the Fermi statistics of electrons. The screening
of the density modulation on scale $v_F \tau_V$ then involves
a time scale $\tau_V$ or longer. As the time resolved noise
correlator also peaks on the time scale $\tau_V$ one
expects that screening modifies the shape of the noise
correlator but preserves its basic form. This conclusion
agrees with the observation that shot noise is usually
observed with the large amplitude obtained within a
non-interacting approximation. Unfortunately, only few detailed
theoretical results are available on the modification of shot
noise due to interaction: in a diffusive conductor the
zero-frequency noise $S(0)$ is even enhanced due to Coulomb
effects \cite{nagaev_95}; the analysis of a ballistic quantum
point contact with a large transmission produces again an
enhancement of $S(0)$, while the shot noise is weakly
reduced in the tunneling limit \cite{zaikin_02}. Finally,
weak interactions can be accounted for via an energy
dependent renormalization of the scattering matrix
\cite{nazarov} leading to a broadening of the electron
wave packets, in agreement with the above discussion.

Dephasing due to interactions among the particles or
with the environment acts differently on the electrons
propagating in the two leads and causes an exponential
damping of the excess correlator on the coherence length
$L_\varphi$. As a consequence, the sum of distances $x_1$
and $x_2$ should be chosen smaller than $L_\varphi$. An
additional requirement is a sufficient experimental time
resolution: with a peak width in $C^\mathrm{ex}_{x_1,x_2}
(\tau)$ given by $\tau_\mathrm{p} = \max[h/eV,1/
\nu_\mathrm{m}]$, with $\nu_\mathrm{m}$ the cutoff
frequency in the measurement setup, only shifts
$|\tau^-| > \tau_\mathrm{p}$ can be resolved. Assuming
a frequency resolution in the 10 GHz regime \cite{schoelkopf_97}
the peak in $C^\mathrm{ex}_{x_1,x_2}(\tau)$ can be resolved
for voltages below 0.1 m$e$V. Given a typical value $v_{\rm
\scriptscriptstyle F}\sim 10^4$ cm/s for the Fermi velocity
this corresponds to a spatial resolution $v_{\rm
\scriptscriptstyle F}/\nu_\mathrm{m} \sim 100$ \AA. The
comparison with a typical mesoscopic dimension of $L\sim
\mu$m scale demonstrates that potential delays expected for
a collapse with a unitary time evolution can be observed
well beyond the scale of the Fermi velocity. However,
the observation of a superluminal collapse would require
frequencies $c/L \sim 10^{14}$ s$^{-1}$ in the 100 THz
regime as well as large voltages of the order of Volts,
both well beyond the acceptable range.

To conclude, we have suggested an experiment testing for
the instantaneous wave function collapse in a solid-state
setup based on the time resolved measurement of
current-current cross correlations at spatially separated
points. This scheme allows to investigate details of the
wave function collapse itself, provided a sufficiently
high frequency resolution is available in the experiment.
While measurements of time delays due to a unitary
collapse involving super-Fermi velocities are within
experimental reach, the type of mesoscopic setup described
here cannot trace time delays arising from a collapse
involving superluminal velocities.

We acknowledge financial support from the Swiss National
Foundation (SCOPES and CTS-ETHZ), the Landau Scholarship
of the FZ J\"ulich, the Russian Science Support Foundation,
the Russian Ministry of Science, and the program `Quantum
Macrophysics' of the RAS.

\end{document}